\documentclass[aps,prd,onecolumn,groupedaddress,showpacs,nofootinbib,amssymb]{revtex4}
\usepackage[dvips]{graphicx}
\usepackage{amssymb}
\usepackage{amsmath}
\usepackage{graphicx,,color}
\usepackage{amsfonts}
\usepackage{bm}
\usepackage{xcolor}
\usepackage{cancel}
\usepackage{comment}
\newcommand\be{\begin{equation}}
\newcommand\ee{\end{equation}}

\DeclareUnicodeCharacter{2212}{-}
\allowdisplaybreaks[4]

\begin{document}
\tolerance=5000

\title{Swampland criteria and constraints on inflation in a $f(R,T)$ gravity theory}

\author{V.K.~Oikonomou,$^{1}$\,\thanks{v.k.oikonomou1979@gmail.com}}
\author{Konstantinos-Rafail Revis,$^{2,1}$\,\thanks{reviskostis@gmail.com}}
\author{Ilias C. Papadimitriou,$^{1}$\,\thanks{elias.papajim@gmail.com,ipapadim@auth.gr}}
\author{Maria-Myrto Pegioudi,$^{1}$\,\thanks{mariamyrto@hotmail.gr,mpegioud@auth.gr}}

\affiliation{$^{1)}$ Department of Physics, Aristotle University
of Thessaloniki, Thessaloniki 54124,
Greece\\
$^{2)}$ Institute for Theoretical Particle Physics and Cosmology (TTK),
RWTH Aachen University, D-52056 Aachen, Germany.\\}

\tolerance=5000
\pacs{04.30.−w, 04.50.Kd, 11.25.-w, 98.80.-k, 98.80.Cq}

\begin{abstract}
{In this paper, we worked in the framework of an inflationary $f(R,T)$ theory, in the presence of a canonical scalar field. More specifically, the $f(R,T)=\gamma R+2\kappa\alpha T$ gravity. The values of the dimensionless parameters $\alpha$ and $\gamma$ are taken to be $\alpha \geq 0$ and $0 < \gamma \leq 1$.
The motivation for that study was the striking similarities between the slow-roll parameters of the inflationary model used in this work and the ones obtained by the rescaled Einstein-Hilbert gravity inflation $f(R)=\alpha R$.
We examined a variety
of potentials to determine if they agree with the current Planck Constraints.
In addition, we checked whether these models satisfy the Swampland Criteria and we specified the
exact region of the parameter space that produces viable results for each model. As we mention in Section IV the inflationary $f(R,T)$ theory used in this work can not produce a positive $n_T$ which implies that the stochastic gravitational wave background will not be detectable.
}
\end{abstract}
\maketitle

\section{Introduction}
In the upcoming years, various groundbreaking experiments are expected to take place. Their results could unveil aspects of the Cosmos that we ignored or could force us to review and amend current theories. Experiments like the LISA space mission \cite{Baker:2019nia,Smith:2019wny} or the DECIGO \cite{Seto:2001qf,Kawamura:2020pcg} are some of the most prominent. The importance of these next-generation observations comes from the fact that observing primordial gravitation waves (GW) is in reality proof that inflation occurred.
This is possible only if the primordial spectrum of tensor perturbations is measured self consistency tests of inflation can be performed by comparing the spectrum of the tensor perturbations with one of the slow-roll parameters. Both of them will be introduced in Section \ref{Theoretical framework}.

The majority of space interferometers can probe scales significantly smaller than $10$ Mpc, so they are in a regime, where Cosmic Microwave Background (CMB) scalar curvature perturbations are clearly non-linear. Excluding the Square Kilometer Array (SKA) \cite{Bull:2018lat} and the NANOGrav collaboration \cite{Arzoumanian:2020vkk,Pol:2020igl}, which could possibly test primordial GWs during the last stages of the radiation domination (RD) era, and the Planck collaboration \cite{Akrami:2018odb} which tests large scales of very small frequencies, the remaining of the probes and observations could obtain information from GWs deeply during the RD era, when the universe is opaque and it's impossible to gain information via photons observations. B-mode polarization modes are essential since they indicate that primordial tensor perturbations exist in large scales (multipoles of CMB that are approximately $l\leq10$) and there is a conversion of $E$-modes to $B$-modes, obviously at later times via direct gravitational lensing conversion, but this time for small angular scales in the CMB \cite{Denissenya:2018mqs}.

The theory of inflation, which was developed by Alan Guth \cite{Guth:1980zm}, is usually referred to as the rapidly accelerated era of the Universe, which emerged right after the quantum gravity era. Inflation provided the solution to some of the most puzzling Cosmological problems like the horizon problem, the flatness problem, the problem of initial energy, and the absence of primordial relics. The standard inflation model assumes the existence of a canonical scalar field named inflaton \cite{Linde:2007fr,Gorbunov:2011zzc,Lyth:1998xn,Martin:2018ycu}. The idea of a canonical scalar field is an important key aspect of the most successful theories in particle physics and general in the field of high energy physics. Unfortunately, the only fundamental scalar field observed is the Higgs particle.

Since inflation took place right after the quantum gravity era it
is rational to believe that some marks should be left in the
effective Lagrangian during the inflation era and thus a
modification of the Einstein-Hilbert action should possibly be
applied with non-linear curvature corrections or with a general
modification of the used action. Modified gravity theory has a lot
of different forms and is considered as one of the most prominent
candidate class of theories which contains higher order curvature
corrections
\cite{reviews1,reviews2,reviews3,reviews4,reviews5,reviews6}. A
large variety of works have unified the late-time accelerated era
with the inflationary one, see fruitful works like
\cite{Nojiri:2003ft} and Refs.
\cite{Nojiri:2007as,Nojiri:2007cq,Cognola:2007zu,Nojiri:2006gh,Appleby:2007vb,Elizalde:2010ts,Odintsov:2020nwm,Oikonomou:2020qah,Oikonomou:2020oex}.
One of the approaches that try to solve these problems is an
$f(R,T)$ gravity proposed by Harko et al. in
Ref.\cite{Harko:2011kv}. In this work, the $f(R,T)$ gravity theory
proposed by M.Gamonal in Ref.\cite{Gamonal:2020itt}, slightly
alternated, will be studied. Then the effective potentials studied
in Ref.\cite{Oikonomou:2021zfl} shall be considered for the
Lagrangian and we will obtain further constraints for the
introduced free parameters. In the next section, the second key
aspect of the presented work will be introduced, which are the
Swampland Criteria  (SC). The final goal of this paper is to
obtain the constraints for the introduced free parameters, imposed
by SC and finally deduce if it is possible to have an agreement
between them.

In this work the sections are structured as follows: {In section
II we present the Swampland Criteria for an inflationary theory,
and in section III we specify the slow-roll conditions and
observational parameters for the specific $f(R,T)$ model that we
studied. Section IV contains the Planck 2018 results as well as
the methodology we followed in this paper. In section V we study
several potentials and derive their constraints in light of the
Planck results and the Swampland criteria. Finally, the
conclusions and bibliography follow at the end of the article. }

\section{Swampland Criteria for rescaled gravity}
In this section, the SC for a gravity theory will be discussed. Their concept was first introduced in the Refs \cite{Vafa:2005ui,Ooguri:2006in} and they have been studied in detail in Refs. \cite{Palti:2020qlc,Mizuno:2019bxy,Brandenberger:2020oav,Blumenhagen:2019vgj,Wang:2019eym,Benetti:2019smr,Palti:2019pca,Cai:2018ebs,Akrami:2018ylq,Mizuno:2019pcm,Aragam:2019khr,Brahma:2019mdd,Mukhopadhyay:2019cai,Yi:2018dhl,Gashti:2022hey,Brahma:2019kch,Haque:2019prw,Heckman:2019dsj,Acharya:2018deu,Elizalde:2018dvw,Cheong:2018udx,Heckman:2018mxl,Kinney:2018nny,Garg:2018reu,Lin:2018rnx,Park:2018fuj,Olguin-Tejo:2018pfq,Fukuda:2018haz,Wang:2018kly,Ooguri:2018wrx,Matsui:2018xwa,Obied:2018sgi,Agrawal:2018own,Murayama:2018lie,Marsh:2018kub,Storm:2020gtv,Trivedi:2020wxf,Sharma:2020wba,Odintsov:2020zkl,Mohammadi:2020twg,Trivedi:2020xlh,Oikonomou:2021zfl}. These are criteria that are imposing conditions, which propose if an effective field theory is or not the correct description for quantum gravity for the high energy scales. So, an effective theory must satisfy the upcoming conditions:

\begin{itemize}
  \item The Swampland distance conjecture, which limits the validity of an effective field theory by setting an upper bound for the maximum traversable range for a scalar field as following:
  \begin{equation}
  \centering
  \label{criterion1}
  |\kappa\Delta\phi|<\mathcal{O}(1)\, ,
  \end{equation}
  \item The de Sitter conjecture, which sets a lower bound for a scalar potential that is positive and its first derivative with respect to the scalar field. In reality, this states that it is impossible to create De Sitter vacua in string theory and reads as:
  \begin{equation}
  \centering
  \label{criterion2}
  \frac{|V'(\phi_i)|}{\kappa V(\phi_i)}>\mathcal{O}(1).
  \end{equation}
   We also have an interchangeable condition which includes the second derivative of the potential with respect to the field, which reads as follows:
  \begin{equation}
  \centering
  \label{criterion3}
  -\frac{V''(\phi_i)}{\kappa^2V(\phi_i)}>\mathcal{O}(1)\, ,
  \end{equation}
\end{itemize}
note that the prime represents the differentiation with respect to the scalar field $\phi$, $\kappa^2=\frac{1}{M_P^2}$ and $M_P$ is the reduced Planck mass and $\phi_i$ denotes the value of the scalar field during the first horizon crossing. At this point is important to underline that we check which values of the free parameters of the potentials introduced in Ref.\cite{Akrami:2018odb} should be satisfied so that the SC are fulfilled. The reader should keep in mind that if at least one of the conditions \ref{criterion1}, \ref{criterion2} or \ref{criterion3} are satisfied the SC are met.

\section{$f(R,T)$ gravity, slow-roll conditions and observational parameters}\label{Theoretical framework}

In this paper, it will not be provided an extensive presentation of the slow-roll parameters and the derivation of the observational indices since these have been presented exceptionally in Ref.\cite{Gamonal:2020itt}. Therefore it would be beneficial to directly continue with the analysis of the slow-roll conditions and observational parameters for a $f(R,T)$ gravity. Firstly let us introduce the definition of follow-momentum tensor $T_{\mu\nu}$:
\begin{equation}
  \centering
  \label{EMT_definition}
   T_{\mu\nu}\equiv-\frac{2}{\sqrt{-g}}\frac{\delta(\sqrt{-g}\mathcal{L}_m)}{\delta g^{\mu\nu}}=g_{\mu\nu}\mathcal{L}_m -2\frac{\delta\mathcal{L}_m}{\delta g^{\mu\nu}} ,
  \end{equation}

The background metric is a flat Friedmann-Robertson-Walker (FRW) metric, which has the following form:

\begin{equation}
\centering
\label{metric}
ds^2=-dt^2+a^2(t)\delta_{ij}dx^idx^j,\,
\end{equation}
where $a(t)$ corresponds to the scale factor. As it is mentioned in the introduction the simplest scenario for inflation assumes the existence of the inflaton $\phi=\phi(t)$. The slow roll conditions were introduced to ensure that the inflationary era lasted long enough to solve the problems presented in the introduction. The slow-roll conditions can be quantified thanks to the slow-roll parameters and in this paper, an extensive presentation of the slow-roll parameters and the derivation of the observational indices will not be provided since this has already been presented in Ref.\cite{Hwang:2005hb}, so only the most relevant for this work will be provided:

\begin{equation}
\centering
\label{epsilon1}
\epsilon_1=-\frac{\dot H}{H^2},
\end{equation}

\begin{equation}
\centering
\label{epsilon2}
\epsilon_2=\frac{\ddot\phi}{H\dot\phi},
\end{equation}

At this point we have to turn our focus on the the definition of the parameters $\epsilon$ and $\eta$, which are given as follows:

\begin{equation}
\centering \label{epsilon}
\epsilon=\frac{1}{2 \kappa^2}\frac{V'^2}{V^2},
\end{equation}

\begin{equation}
\centering
\label{eta}
\eta=\frac{1}{\kappa^2}\frac{V''}{V}.
\end{equation}
It essential for our analysis to introduce,
\begin{equation}
\centering \label{epsilonV}
\epsilon_V=\epsilon_1,
\end{equation}

\begin{equation}
\centering \label{etaV}
\eta_V=\epsilon_1-\epsilon_2.
\end{equation}
At this point, it is extremely important to avoid a very common confusion. For the classical Einstein-Hilbert action the equation (\ref{epsilon1}) leads directly to the equation (\ref{epsilon}) and vice versa. But for a general $f(R,T)$ gravity theory this is not the case even if it is obvious to find a correspondence between the $f(R,T)$ gravity theory and the classical one which is obtained by the Einstein-Hilbert action. At this point, we have also to underline the connection between the SC provided by equation \ref{criterion1} and the $\epsilon$ provided by equation (\ref{epsilon}), for $\kappa=1$ reads,
\begin{equation}
    \centering\label{S.C. with epsilon}
    \Big{|}\frac{V'}{V} \Big{|}=\sqrt{2 \epsilon}.
\end{equation}

The reader could possibly assume the existence of a tension between the satisfaction of the slow-roll conditions imposed to the slow-roll parameter and the condition for the satisfaction of the SCs. This obstacle is possible to overcome if a vital detail is recalled. The slow-roll indices are described
by the conditions  $\epsilon_1\ll1$ and $\epsilon_2\ll1$ and not
by the conditions $ \epsilon\ll 1$ and $\eta\ll 1$. This is an important detail, which should be taken into consideration to avoid any confusion. To make these results crystal-clear we need to introduce the action, derive the equations of motion (EoM) and a rescaled Klein-Gordon (KG) equation. The general formulation of the EoM for an $f(R,T)$ gravity theory is presented in the Ref.\cite{Gamonal:2020itt} and since this work is limited in a $f(R,T)=\gamma R+2\kappa\alpha T$ it is beneficial the action to be directly provided. So, for $\kappa=1$, we have that:

\begin{equation}
\centering\label{action}
\mathcal{S}=\int{d^4x\sqrt{-g}\left( \frac{\gamma R}{2\kappa^2} + \alpha T +\mathcal{L}_m
\right)},
\end{equation}
so we can arrive at the equations of motion which according to the Ref.\cite{Gamonal:2020itt} read:

\begin{equation}
\centering\label{eom}
\gamma R_{\mu\nu}-\frac{1}{2}g_{\mu\nu}\gamma R= T_{\mu\nu} - 2\alpha\left(T_{\mu\nu}-\frac{1}{2}T g_{\mu\nu} +\Theta_{\mu\nu}
\right),
\end{equation}
The action in Eq. (\ref{action}) may be the result of some higher
order curvature effects at leading order during the inflationary
era. For example  \cite{Oikonomou:2020oex},
\begin{equation}\label{frini}
f(R)=R-\alpha  \lambda  \Lambda -\lambda  R \exp
\left(-\frac{\alpha  \Lambda }{R}\right)-\frac{\Lambda
\left(\frac{R}{m_s^2}\right)^{\delta }}{\zeta }\, .
\end{equation}
In the large curvature limit, the exponential term of Eq.
(\ref{frini}) at leading order yields,
\begin{equation}\label{expapprox}
\lambda  R \exp \left(-\frac{\gamma  \Lambda }{R}\right)\simeq
-\alpha \lambda  \Lambda -\frac{\alpha ^3 \lambda \Lambda^3}{6
R^2}+\frac{\alpha ^2 \lambda  \Lambda ^2}{2 R}+\lambda  R\, ,
\end{equation}
hence, the effective action  during inflation contains terms of
the Ricci scalar as follows,
\begin{equation}\label{effectiveaction}
\mathcal{S}=\int
d^4x\sqrt{-g}\left(\frac{1}{2\kappa^2}\left(\gamma R+ \frac{\alpha
^3 \lambda \Lambda ^3}{6 R^2}-\frac{\alpha ^2 \lambda \Lambda
^2}{2 R}-\frac{\Lambda}{\zeta
}\left(\frac{R}{m_s^2}\right)^{\delta
}+\mathcal{O}(1/R^3)+...\right)-\frac{1}{2}g^{\mu\nu}\nabla_\mu\phi\nabla_\nu\phi-V(\phi)-\xi(\phi)\mathcal{G}\right)\,
,
\end{equation}
where $\gamma=1-\lambda$. So after the analysis we arrive at the
following equations:

\begin{equation}
\centering\label{eom1}
H^2= \frac{\kappa^2}{3 \gamma}\left(\frac{\dot\phi^2}{2}(1+2\alpha) +V(\phi)(1+4\alpha)
\right),
\end{equation}

\begin{equation}
\centering\label{eom2}
\frac{\ddot\alpha}{\alpha}=-\frac{\kappa^2}{3\gamma}\left(\dot\phi^2(1+2\alpha) -V(\phi)(1+4\alpha)
\right),
\end{equation}

\begin{equation}
\centering\label{eom3}
\dot H  = \frac{\ddot\alpha}{\alpha} -H^2 =-\frac{\kappa^2\dot\phi^2}{2\gamma}\left(1+2\alpha
\right).
\end{equation}

Finally the rescaled KG equation can be obtained, which according to Ref.\cite{Gamonal:2020itt} reads as:

\begin{equation}
\centering\label{rescaledKG}
\ddot\phi(1+2\alpha) + 3H\dot\phi(1+2\alpha) + \frac{\partial V }{\partial\phi}(1+4\alpha) = 0.
\end{equation}

It should be mentioned that all the previous results reproduce the well-know results for the classical Einstein-Hilbert action by simply setting $\alpha=0$. At this point the spectral indices and their connection to the slow-roll parameters should be introduced, which according to the Ref.\cite{Gamonal:2020itt} read as:

\begin{equation}
\centering\label{ns}
n_s-1=\frac{dln(\Delta^2_s)}{dln(k)}=-4\epsilon_1+2\epsilon_2,
\end{equation}

\begin{equation}
\centering\label{nt}
n_T=\frac{dln(\Delta^2_T)}{dln(k)}=-2\epsilon_1,
\end{equation}

\begin{equation}
\centering\label{r}
r=\frac{\Delta^2_T}{\Delta^2_s}=16\epsilon_1.
\end{equation}
It should be mentioned, that up to this date there is no value for
$n_T$ since the B-mode polarization for GW have not been observed.
Recalling the definition of the $\epsilon_1$ by equation
(\ref{epsilon1}) for the model one can obtain:

\begin{equation}
\centering\label{our_epsilon1}
\epsilon_1=\frac{3}{2}\left(\frac{\dot\phi^2(1+2\alpha)}{\frac{\dot\phi^2}{2}(1+2\alpha)+V(\phi)(1+4\alpha)}\right).
\end{equation}
 The slow-roll parameters are formulated in such a way that we are able to quantify directly the required conditions to impose slow-roll conditions. For $\epsilon_1$ we demand that $\epsilon_1\ll1$. Therefore the condition is:

\begin{equation}
\centering\label{SR-condition1}
\dot\phi^2(1+2\alpha)\ll V(\phi)(1+4\alpha),
\end{equation}
so $\epsilon_1$ is now approximately equal to,

\begin{equation}
\centering\label{our_epsilon1+SR}
\epsilon_1 \approx\frac{3(1+2\alpha)}{2(1+4\alpha)}\frac{\dot\phi^2}{V(\phi)}.
\end{equation}
The KG becomes:

\begin{equation}
\centering\label{rescaledKG+SR}
 3H\dot\phi(1+2\alpha) + \frac{\partial V }{\partial\phi}(1+4\alpha) \approx 0,
\end{equation}
while equation (\ref{eom1}) take the following form,
\begin{equation}
\centering\label{eom1+SR}
H^2 \approx \frac{\kappa^2 (1+4\alpha)}{3 \gamma}V(\phi).
\end{equation}

Following the methodology introduced in the Ref.\cite{Gamonal:2020itt}, we can arrive at the derivation of the the observational indices with respect to the potential and the free parameters $\alpha$ and $\gamma$ introduced by our $f(R,T)$ gravity model. We obtain the final expression for $\epsilon_1$ by substituting $\dot{\phi}$ from equation (\ref{rescaledKG+SR}) and $H$ from equation (\ref{eom1+SR}) in equation (\ref{our_epsilon1+SR}).  So the primordial tilt reads as:

\begin{equation}
\centering\label{ns_calc}
 n_s=1+2\eta_V-6\epsilon_V,
\end{equation}

\begin{equation}
\centering\label{nt_calc}
 n_T=-2\epsilon_V,
\end{equation}

\begin{equation}
\centering\label{r_calc}
 r=16\epsilon_V,
\end{equation}
where
\begin{equation}
\centering\label{epsilon_v}
\epsilon_V=\left(\frac{\gamma}{1+2\alpha}\right)\epsilon,
\end{equation}
and
\begin{equation}
\centering\label{eta_v}
 \eta_V=\left(\frac{\gamma}{1+2\alpha}\right)\eta.
\end{equation}
Here we come to an interesting realization, the slow roll indices and by extension $n_s$, $n_T$ and $r$ depend only on the value of $\frac{\gamma}{1+2\alpha}$, which as we will soon see holds true for the e-folds number too. For that reason it is convenient to use $\beta(\gamma, \alpha)=\frac{\gamma}{1+2\alpha}$ in order to simplify the analysis and the rather complicated expressions that we will come across in this work.
\section{Planck 2018 results and methodology presentation}\label{methodology}
The goal of this section is to present the methodology that will be followed in this work. Firstly the value of the inflaton at the end of inflation, $\phi=\phi_{end}$ is obtained by using the equation $\epsilon_V(\phi_{end})=1$. At this point, the definition of the e-folding number should be recalled $N\equiv \ln\alpha$, so after some simple calculations we arrive at:

\begin{equation}
\centering\label{N_calc}
 N\approx\kappa^2\frac{1+2\alpha}{\gamma}\int_{\phi_{end}}^{\phi_{i}}{\frac{V}{V'}d\phi},
\end{equation}
where $V'=\frac{\partial V }{\partial\phi}$. So, after calculating the number of e-folds, the $\phi_i$ can be obtained, which is essential to calculate since $n_s$, $n_T$ and $r$
are calculated for $\phi=\phi_i$. Then using the equations (\ref{ns_calc}), (\ref{nt_calc}) and (\ref{r_calc}) the observational indices are obtained as a function of $N$, $\alpha$ and possibly the free parameters the introduced potential has. It should be underlined that only two free parameters should be left in the final obtained results for the spectral indices so if there are more than three left the e-folding number will be eliminated by setting $N=60$. Then, the remaining parameters are constraint using the Planck 2018 results from Ref.\cite{Akrami:2018odb}. Namely,
\begin{equation}\label{PlankConstraints}
    n_s=0.9649 \pm 0.0042 \ (68\% \ CL) \ , \ r<0.056 \ (95 \%  \ CL).
\end{equation}
We should mention that the parameter $\beta(\gamma, \alpha)$ can only take positive values because a negative $\beta(\gamma, \alpha)$ would result in a negative $r$ as can be seen in the expressions (\ref{r_calc}) and (\ref{epsilon_v}). Therefore, in this $f(R,T)$ theory $n_T$  can only take negative values, which means that the primordial gravitational waves produced will not be detectable by any of the upcoming gravitational wave experiments. \\
At this point we turn our focus on the on the SC. By using the equations (\ref{criterion1}), (\ref{criterion2}) and (\ref{criterion3}) we are able to impose new constraints on the introduced free parameters and then compare with the ones we found on the first part of our analysis and finally deduce if it possible that SC and inflationary constraints are satisfied at the same time for the model we are studying.

\section{Derivation of the constraints on the inflationary models and SC.}\label{analysis}

\subsection{Power law potentials}\label{Power Laws}
Let us start with the study of the general family of the power laws potentials. In general it can be written:
\begin{equation}
    \centering\label{v_power laws}
    V(\phi)=\Lambda(\phi \kappa )^p,
\end{equation}
where $\Lambda$ has $4-p$ mass dimensions. Since the methodology used was presented in detail in  Section \ref{methodology} the obtained results can be directly presented. The slow roll parameters for $\kappa=1$ read:

\begin{equation}
    \centering\label{eps_power laws}
    \epsilon_V=\frac{\beta(\gamma,\alpha)  p^2}{2 \phi ^2}
\end{equation}

\begin{equation}
    \centering\label{eta_power laws}
  \eta_V=\frac{\beta(\gamma,\alpha)  (p-1) p}{\phi ^2}.
\end{equation}
At this point we can proceed for $\kappa=1$ and arrive at the calculation of $\phi_i$:

\begin{equation}
    \centering\label{phi_i_power laws}
  \phi_i=\frac{\sqrt{p \beta(\gamma,\alpha)}\sqrt{4N+p}}{\sqrt{2}}.
\end{equation}
So, we are now able to derive the spectral indices for $\phi=\phi_i$:

\begin{equation}
    \centering\label{ns_power laws}
  n_s=1-\frac{2 (2+p)}{p+4  N}
\end{equation}

\begin{equation}
    \centering\label{r_power laws}
  r=\frac{16 p}{p+4 N}.
\end{equation}
This is the most important part of the presented work since there are many ways to obtain the required bounds for the free parameters. It was chosen to eliminate $\beta$ since $\beta=\frac{\gamma}{2 \alpha +1}=\beta(\gamma,\alpha)$, $\kappa=1$ was setted and and when it was necessary to end up with only 2 parameters free we set $N=60$.

\begin{figure}[h!]
\centering
\includegraphics[width=18pc]{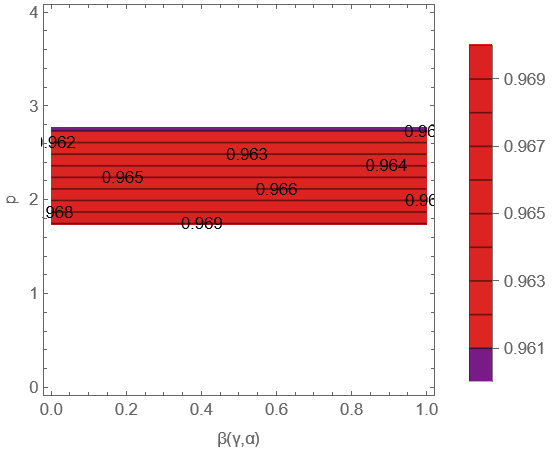}
\includegraphics[width=18pc]{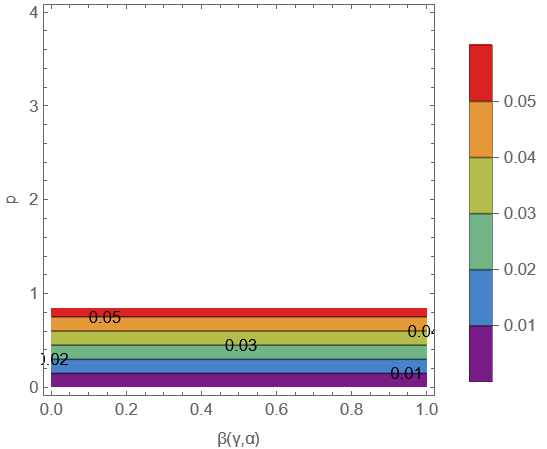}
\caption{Contour plot for the spectral index of primordial scalar
curvature perturbations $n_s$ (left plot) and the tensor-to-scalar
ratio $r$ (right plot) for $\beta(\gamma,\alpha)=[0, 1]$, $p=[0,
4]$ and $N= 60$ for the Power Law potentials. }\label{PL_INFL_fig}
\end{figure}

{
Using the Plank Constraints (\ref{PlankConstraints}) we derived the constraints for the parameter $p$ of the model. In specific, the relation (\ref{ns_power laws}) for the scalar spectral index results in the following inequality,
\begin{equation}
    \centering\label{PL_NS}
    1.7348\leq p\leq 2.77044,
\end{equation}
while from (\ref{r_power laws}) we calculate the inequality,
\begin{equation}
    \centering\label{PL_R}
    0\leq p\leq 0.84295.
\end{equation}
As can be seen, the two inequalities are incompatible, so the model is not viable.
}

Let us proceed to the calculation of the constraints that emerged from the SC. Using the equations (\ref{criterion1}), (\ref{criterion2}) and (\ref{criterion3}) and setting $N=60$ and $\kappa=1$ we arrive at:

\begin{equation}
    \centering\label{PL_sc1}
    \Delta\phi= \frac{\sqrt{p \beta(\gamma,\alpha)}(\sqrt{p}-\sqrt{240+p})}{\sqrt{2}},
\end{equation}

\begin{equation}
    \centering\label{PL_sc2}
   \frac{|V'(\phi_i)|}{\kappa V(\phi_i)} = \frac{2 p}{ \sqrt{(240+p)\beta(\gamma,\alpha)}},
\end{equation}

\begin{equation}
    \centering\label{PL_sc3}
   -\frac{V''(\phi_i)}{\kappa^2 V(\phi_i)} = \frac{2 (1-p)}{\beta(\gamma,\alpha)(240+p)} ,
\end{equation}

\begin{figure}
\centering
\includegraphics[width=18pc]{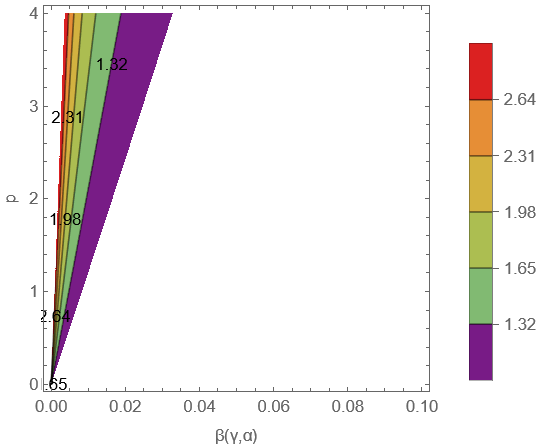}
\includegraphics[width=18pc]{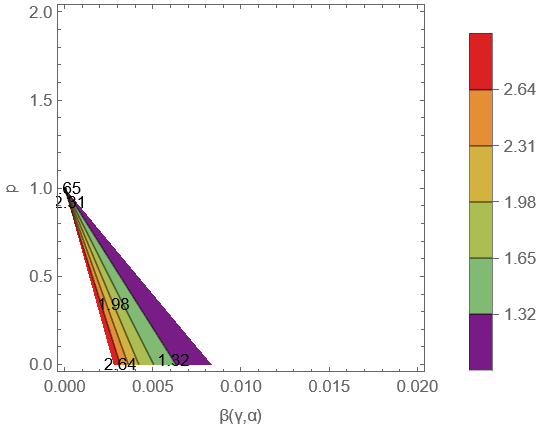}
\caption{Contour plot for de Sitter conjecture. The left plot provides $V'(\phi_i)/V(\phi_i)$ and the right plot provides the $-V''(\phi_i)/V(\phi_i)$. For both cases, the constraints from SC are taken into account. Note that $\beta(\gamma,\alpha)=[0, 0.4]$, $p=[0, 4]$ and $N= 60$ for Power Law potentials.}\label{PL_sc_desitter_fig}
\end{figure}

\begin{figure}
\centering
\includegraphics[width=18pc]{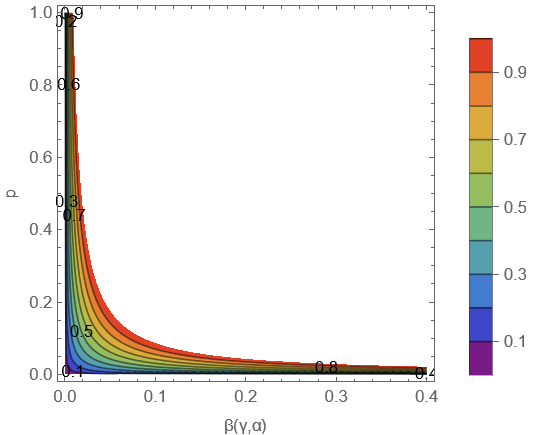}
\caption{Contour plot for the Swampland distance conjecture. Note that $\beta(\gamma,\alpha)=[0, 0.4]$, $p=[0, 4]$ and $N= 60$ for Power Law potentials.}\label{PL_df_fig}
\end{figure}

Taking the equations (\ref{PL_sc1}), (\ref{PL_sc2}) and (\ref{PL_sc3}) we will arrive at the figures Fig.\ref{PL_sc_desitter_fig} and Fig.\ref{PL_df_fig}.

{Using the (\ref{criterion1}) we calculate $p>0$ as well as,
\begin{equation}
    \centering\label{Cr1-PL} \beta(\gamma,\alpha)   \leq \frac{120+p+\sqrt{p(240+p)}}{1440p}.
\end{equation}
For (\ref{criterion2}) to be satisfied,
\begin{equation}
    \centering\label{Cr2-PL}
   \beta(\gamma,\alpha)   \leq  \frac{2p}{240+p}.
\end{equation}
Considering (\ref{criterion3}), we calculate the,
\begin{equation}
    \centering\label{Cr3-PL}
   p\leq \frac{2-240 \beta(\gamma,\alpha)}{2+\beta(\gamma,\alpha)}.
\end{equation}
}

Taking the equations (\ref{PL_sc1}), (\ref{PL_sc2}) and (\ref{PL_sc3}) we will arrive at the figures Fig.\ref{PL_sc_desitter_fig} and Fig.\ref{PL_df_fig}, which prove the existence of a large variety of different values of the introduced free parameters that satisfy the SC. Since the primary goal was to obtain a result combining the two different sets of constraints and in this case this is impossible since the inflationary ones fail and thus no further progress can be achieved. Note that it is known that the Power Law models are fragile to be non-viable for the classic Einstein-Hilbert action and as it is proven in \cite{Oikonomou:2021zfl} this is the situation even for a $f(R)$ gravity approach. So, we decided to present extensively the results in between before we arrive at the final form of the observational indices. We will not insist on presenting extensively the results in between since the methodology we used is precisely given in Section \ref{methodology}.

\subsection{D-Brane p=4}\label{D4}
Lets us proceed by working on the D-Brane model \cite{Akrami:2018odb}:
\begin{equation}\label{D2}
V(\phi)=\Lambda ^4 \left(1-\left(\frac{m}{\kappa\phi}\right)^4\right),
\end{equation}
where $\Lambda$ has dimensions of mass [m] and $m$ is constructed in such a way it is dimensionless. Since the methodology used was presented in detail in  Section \ref{methodology} we can directly proceed with the obtained results.

\begin{equation}
    \centering\label{D4fi}
    \phi_i=\frac{\sqrt[6]{2} \sqrt[6]{2^{4/5} \beta^{3/5}(\gamma,\alpha) m^{24/5}+12 \beta(\gamma,\alpha) m^4 N}}{\kappa}.
\end{equation}
So using the methodology presented in Section \ref{methodology} and setting $\kappa=1$ we arrived at:

\begin{multline*}
 n_s=1-\frac{48 \beta(\gamma,\alpha)m^8}{ \left[-m^4\left(2\cdot 2^{4/5} \beta(\gamma,\alpha)^{3/5} m^{24/5}+1440 m^4 \beta(\gamma,\alpha)\right)^{1/6}+ \left(2 \cdot 2^{4/5} \beta(\gamma,\alpha)^{3/5} m^{24/5}+1440 m^4 \beta(\gamma,\alpha)\right)^{5/6}\right]^2} \\
-\frac{40\beta(\gamma,\alpha)}{\left[2\cdot 2^{4/5} \beta(\gamma,\alpha)^{3/5} m^{4/5}+1440 \beta(\gamma,\alpha)-\left(2 \cdot 2^{4/5} \beta(\gamma,\alpha)^{3/5} m^{24/5}+1440 m^4 \beta(\gamma,\alpha)\right)^{1/3}\right]^2}
\end{multline*}

\begin{equation}
    \centering\label{D4r}
    r=\frac{128 m^8 \beta(\gamma,\alpha)}{\left[-m^{4}\left(2\cdot 2^{4/5} \beta(\gamma,\alpha)^{3/5} m^{24/5}+1440m^4 \beta(\gamma,\alpha)\right)^{1/6}+\left(2\cdot 2^{4/5} \beta(\gamma,\alpha)^{3/5} m^{24/5}+1440 m^4\beta(\gamma,\alpha)\right)^{5/6}\right]^2}\,
    .
\end{equation}

\begin{figure}
\centering
\includegraphics[width=18pc]{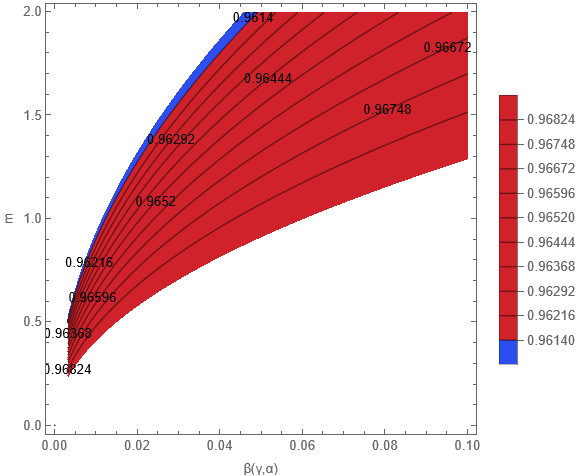}
\includegraphics[width=18pc]{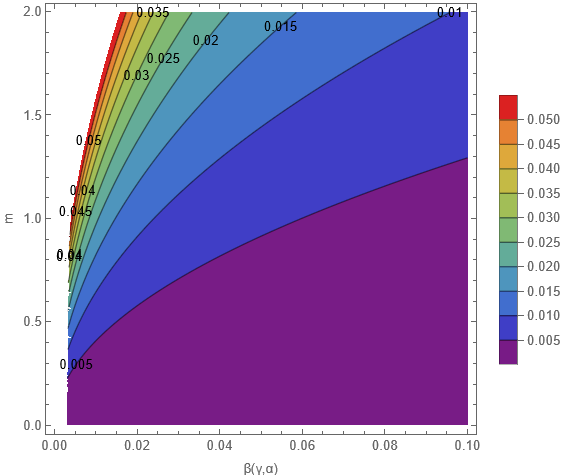}
\caption{Contour plot for the spectral index of primordial scalar curvature perturbations $n_s$ (left plot) and the tensor-to-scalar ratio $r$ (right plot) for $\beta=[0, 0.1]$, $m=[0.000001, 1.995260]$ and $N= 60$ for the D-Brane Model (p=4)}\label{D4_INFL_fig}
\end{figure}

Using the above equations the demanded plots can be obtained to obtain the constraints, which are presented in Fig.\ref{D4_INFL_fig} and it makes obvious that there is indeed an area that constraints the free parameters in such a way that the observational conditions are met. {
Using (\ref{PlankConstraints}) for $n_S$ we found that,
\begin{equation}
    \centering\label{NS_DB}
    16.5113 \leq \frac{m^2}{\beta(\gamma,\alpha)}\leq 87.1122.
\end{equation}
As for the tensor-to-scalar ratio, from (\ref{PlankConstraints}) it follows that,
\begin{equation}
    \centering\label{RS_DB}
    0\leq \frac{m^2}{\beta(\gamma,\alpha)} \leq 251.8.
\end{equation}
From (\ref{NS_DB}) and (\ref{RS_DB}) we get,
\begin{equation}
    \centering\label{NS_AND_R_DB}
    16.5113 \leq \frac{m^2}{\beta(\gamma,\alpha)}\leq 87.1122.
\end{equation}
Thus, the model can be considered viable.
}
Let us proceed to the calculation of the constraints that emerged from the SC. Using the equations (\ref{criterion1}), (\ref{criterion2}) and (\ref{criterion3}) and setting $N=60$ and $\kappa=1$ we arrive at:

\begin{equation}
    \centering\label{d4_sc1}
    \Delta\phi= 2^{3/10} m^{4/5}\beta(\gamma,\alpha)^{1/10} -\sqrt[6]{2\cdot 2^{4/5}m^{24/5} \beta(\gamma,\alpha)^{3/5} +1440 m^4\beta(\gamma,\alpha)} ,
\end{equation}

\begin{equation}
    \centering\label{d4_sc2}
   \frac{|V'(\phi_i)|}{\kappa V(\phi_i)} = -\frac{4 m^4}{\left(m^4 \left(2\cdot 2^{4/5}m^{24/5} \beta(\gamma,\alpha)^{3/5} +1440 m^4\beta(\gamma,\alpha)\right)^{1/6}-\left(2\cdot 2^{4/5}m^{24/5} \beta(\gamma,\alpha)^{3/5} +1440 m^4\beta(\gamma,\alpha)\right)^{5/6}\right)},
\end{equation}

\begin{equation}
    \centering\label{d4_sc3}
  - \frac{V''(\phi_i)}{\kappa^2 V(\phi_i)} = \frac{20}{\left(2\cdot2^{4/5} m^{4/5}\beta(\gamma,\alpha)^{3/5}+1440\beta(\gamma,\alpha)-(2\cdot2^{4/5}m^{24/5}\beta(\gamma,\alpha)^{3/5}+1440m^4\beta(\gamma,\alpha))^{1/3}\right) },
\end{equation}

\begin{figure}
\centering
\includegraphics[width=18pc]{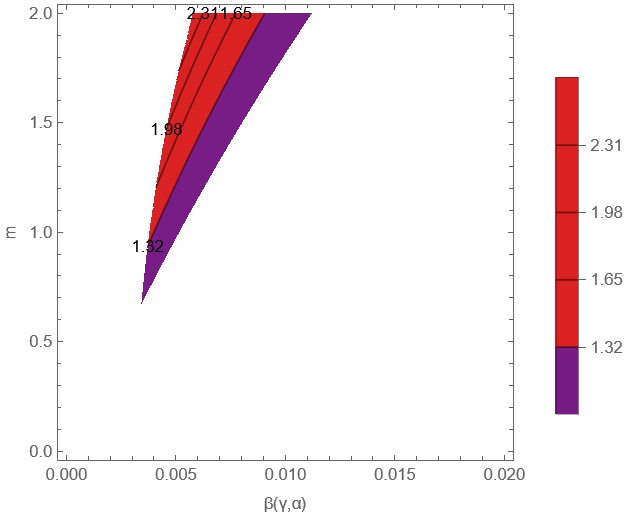}
\includegraphics[width=18pc]{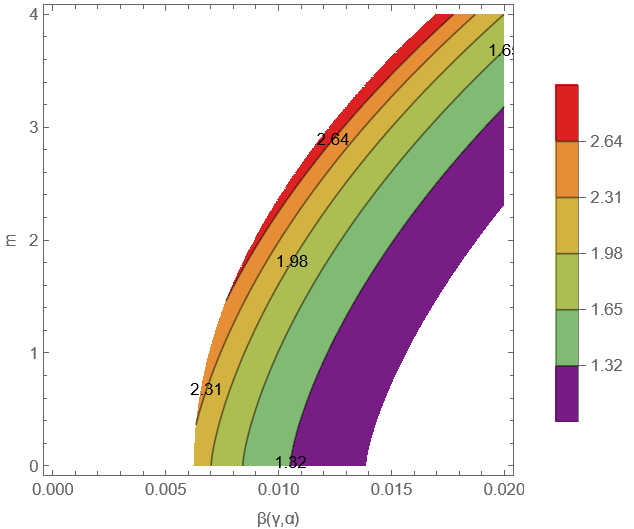}
\caption{Contour plot for de Sitter conjecture. The left plot provides $V'(\phi_i)/V(\phi_i)$ and the right plot provides the $-V''(\phi_i)/V(\phi_i)$. For both cases, the constraints from SC are taken into account. Note that $\beta(\gamma,\alpha)=[0.00, 0.02]$, $m=[0, 2]$ and $m=[0,4]$ and $N= 60$ for the D-Brane Model (p=4)}\label{d4_sc_desitter_fig}
\end{figure}

\begin{figure}
\centering
\includegraphics[width=18pc]{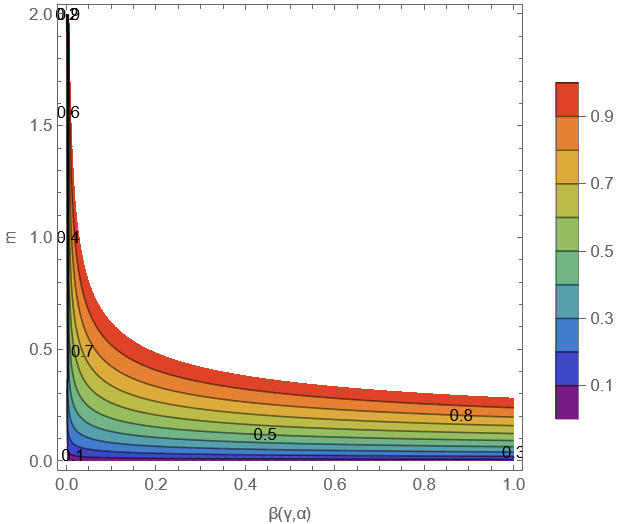}
\caption{Contour plot for the Swampland distance conjecture. Note that $\beta(\gamma,\alpha)=[0, 1]$, $m=[0,2]$ and $N= 60$ for the D-Brane Model (p=4).}\label{d4_df_fig}
\end{figure}

Taking the equations (\ref{d4_sc1}), (\ref{d4_sc2}) and (\ref{d4_sc3}) we will arrive at the figures Fig.\ref{d4_sc_desitter_fig} and Fig.\ref{d4_df_fig}.

Using (\ref{criterion1}), (\ref{criterion2}), (\ref{criterion3}) we examined numerically their behavior for $w \leq 2$ and $\beta(\gamma,\alpha) \leq 0.3$.

For the first SC (\ref{criterion1}) we calculate,
\begin{equation}
\begin{aligned}
    \centering\label{Cr1-D}
   m\leq & 0.392602 + 4.39106 \cdot 10^{-11} \beta^{-5}(\gamma,\alpha) - 2.17079 \cdot 10^{-8} \beta^{-4}(\gamma,\alpha) +
 4.14426 \cdot 10^{-6} \beta^{-3}(\gamma,\alpha) - 3.97486 \cdot 10^{-4} \beta^{-2}(\gamma,\alpha) +\\
 &+ 0.0274175 \beta^{-1}(\gamma,\alpha) - 0.163987 \beta(\gamma,\alpha).
\end{aligned}
\end{equation}
\\For the (\ref{criterion2}) to be satisfied,
\begin{equation}
    \centering\label{Cr2-D}
   m \geq -0.0123206 + 211.331 \beta(\gamma,\alpha) - 2852.67 \beta^{2}(\gamma,\alpha).
\end{equation}
Considering the (\ref{criterion3}), we calculate the
\begin{equation}
    \centering\label{Cr3-D}
   m \geq -9.26722 + 938.092 \beta(\gamma,\alpha) - 18021.4 \beta^2(\gamma,\alpha) .
\end{equation}

\begin{figure}[h]
\centering
\includegraphics[width=18pc]{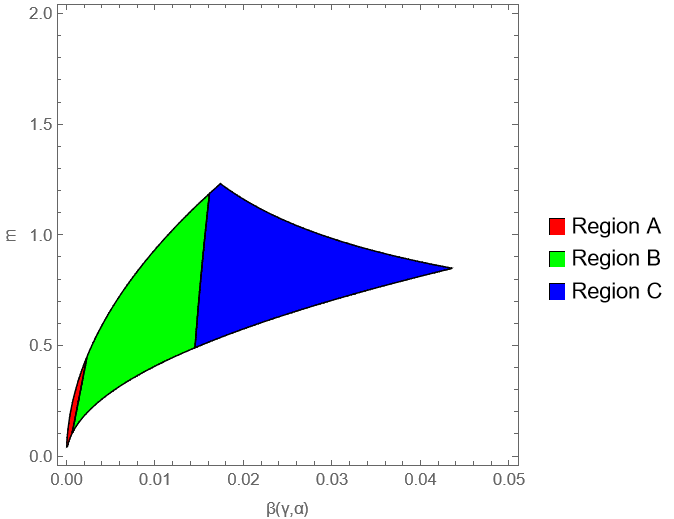}
\caption{Region plot for the combined constraints for D-Brane Model(p=4). In all highlighted Regions the Planck Constraints and at least one Swampland Criterion are satisfied. Region A is the area where all the SC are satisfied, in Region B (\ref{criterion1}) and (\ref{criterion3}) are satisfied simultaneously. In Region C (\ref{criterion1}) is satisfied.}
\end{figure}

\subsection{E model $n=1$}
In this subsection, we proceed with the analysis of rather interesting potential, the E model for $n=1$, which reads as:
\begin{equation}
\centering \label{t1_model_potential}
V(\phi)=\Lambda ^4 \left(1-e^{-\frac{\sqrt{\frac{2}{3}} \kappa \phi}{\sqrt{w}}}\right)^2,
\end{equation}
where $\Lambda$ has mass dimensions and $w$ has no dimensions, in general, we are using the free parameters in such a way they are usually dimensionless because it is just slightly easier to work with. Flowing the methodology presented in section \ref{methodology} we derive the $\phi_i$, which reads as:
\begin{equation}
\centering\label{e1_fini}
\phi_i=\frac{\sqrt{\frac{3}{2}} \sqrt{w} \log \left(\frac{4 \beta(\gamma,\alpha) N}{3 w}+\frac{2 \sqrt{\beta(\gamma,\alpha)}}{\sqrt{3} \sqrt{w}}+1\right)}{\kappa}.
\end{equation}
So using that $\beta(\gamma,\alpha)=\frac{\gamma}{1+2\alpha}$, the methodology presented in Section \ref{methodology} and setting $\kappa=1$ we arrived at:

\begin{equation}
\centering \label{E-1_ns}
n_s=\frac{-9w+236\sqrt{3w\beta(\gamma,\alpha)}+13920\beta(\gamma,\alpha)}{\left(\sqrt{3w}+120\sqrt{\beta(\gamma,\alpha)}\right)^2},
\end{equation}
and
\begin{equation}
\centering \label{E-1_r}
r=\frac{48w}{\left(\sqrt{3w}+120\sqrt{\beta(\gamma,\alpha)}\right)^2}.
\end{equation}

\begin{figure}
\centering
\includegraphics[width=18pc]{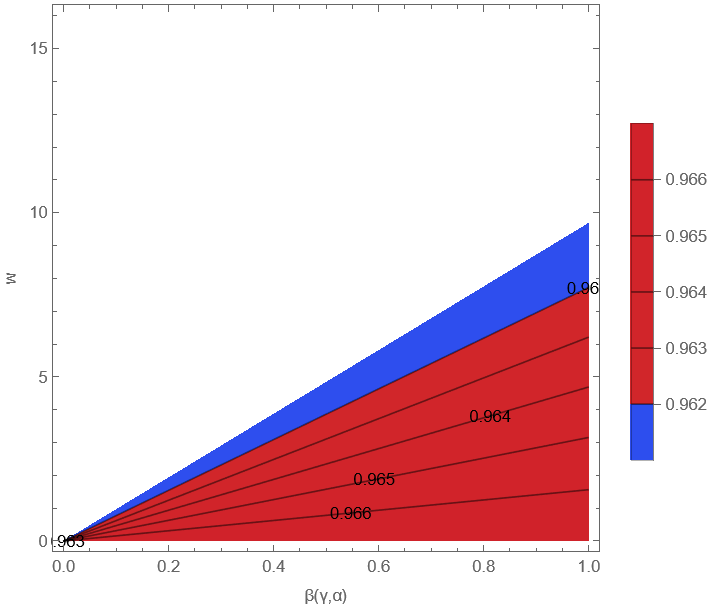}
\includegraphics[width=18pc]{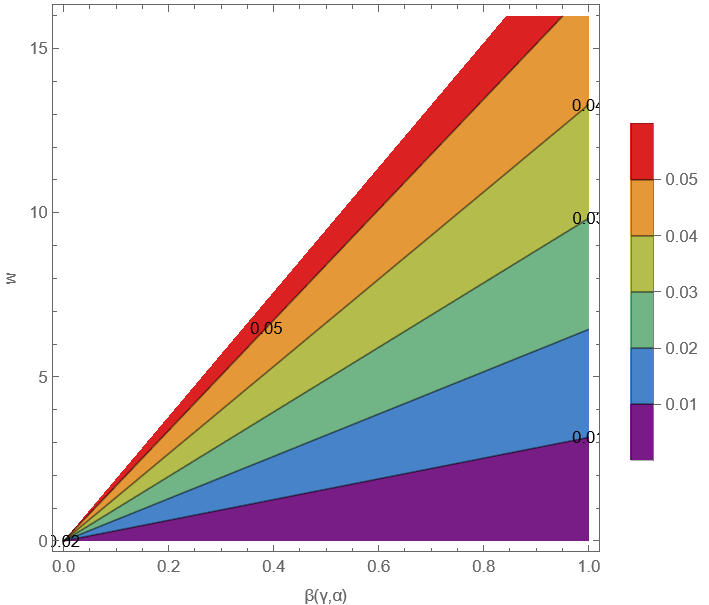}
\caption{Contour plot for the spectral index of primordial scalar curvature perturbations $n_s$ (left plot) and the tensor-to-scalar ratio $r$ (right plot) for $\beta(\gamma,\alpha)=[0, 1]$, $w=[10^{-6}, 16]$ and $N= 60$ for the E-Model (n=1)}\label{E1_INFL_fig}
\end{figure}

Using the above equations we can arrive at the demanded plots to obtain the constraints, which are presented in Fig.\ref{E1_INFL_fig} and it makes obvious that there is indeed an area that constraints the free parameters in such a way that the observational conditions are met. {
Using the Plank Constraints (\ref{PlankConstraints}) we derived the constraints for the parameters $w$ and $\alpha$ of the model. In specific, the relation (\ref{E-1_ns}) for the scalar spectral index results in the following inequality,
\begin{equation}
    \centering\label{E-Pl.Con}
    \frac{w}{\beta(\gamma,\alpha)}\leq 9.69674,
\end{equation}
while from (\ref{r_power laws}) we calculate the inequality,
\begin{equation}
    \centering\label{E-Pl.Conr}
    \frac{w}{\beta(\gamma,\alpha)}\leq 18.9792 \, .
\end{equation}
As a result, we end up with,
\begin{equation}
    \centering\label{E-Combined}
    \frac{w}{\beta(\gamma,\alpha)} \leq 9.69674,
\end{equation}
so the model is viable.}

\begin{figure}
\centering
\includegraphics[width=18pc]{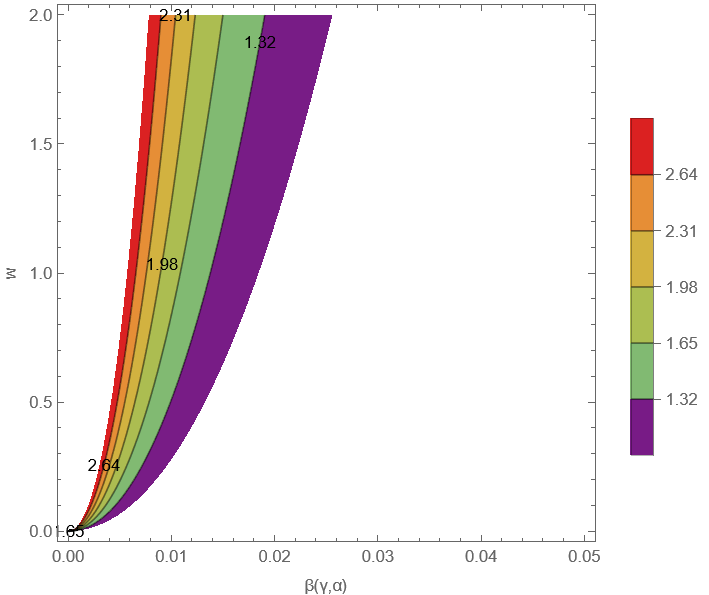}
\includegraphics[width=18pc]{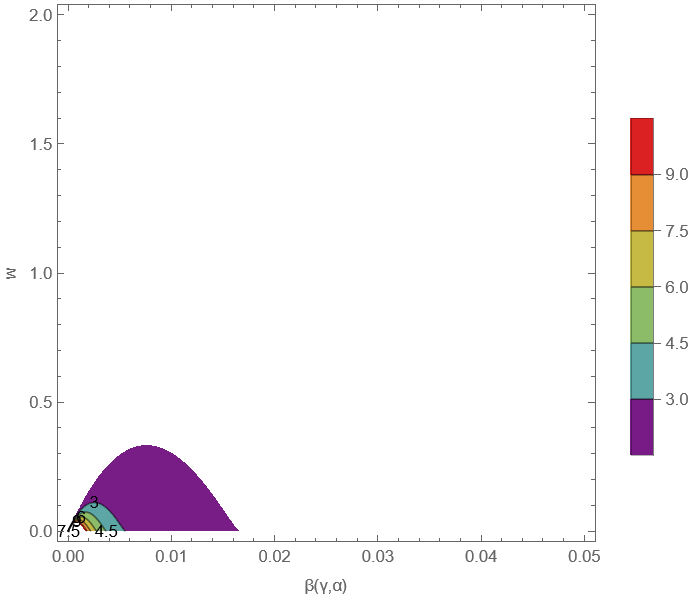}
\caption{Contour plot for de Sitter conjecture. The left plot provides $V'(\phi_i)/V(\phi_i)$ and the right plot provides the $-V''(\phi_i)/V(\phi_i)$. For both cases the constraints from SC are taken into account. Note that $\beta(\gamma,\alpha)=[0, 0.05]$, $w=[0, 2]$ (left plot) and $\beta(\gamma,\alpha)=[0, 0.05]$, $w=[0, 2]$ (left plot) and $N= 60$ for the E-model (n=1).}\label{E1_sc_desitter_fig}
\end{figure}

\begin{figure}
\centering
\includegraphics[width=18pc]{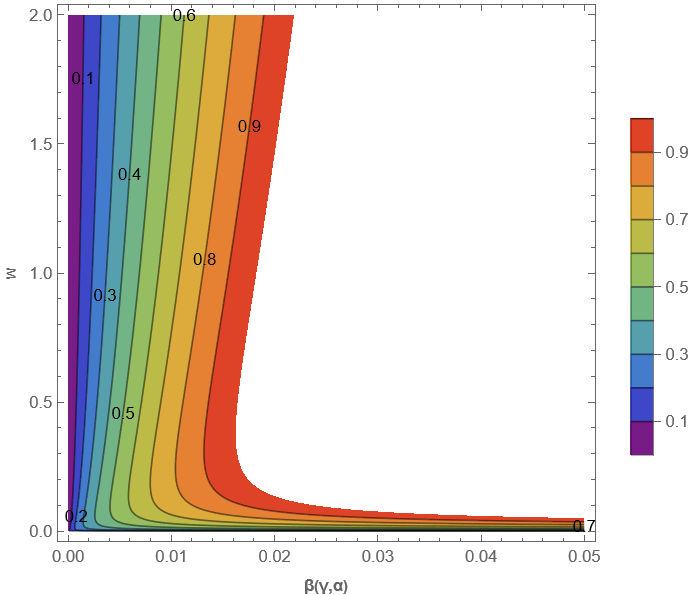}
\caption{Contour plot for the Swampland distance conjecture.  Note that $\beta(\gamma,\alpha)=[0, 0.05]$, $w=[0, 2]$ and $N= 60$ for the E-model (n=1).}\label{E1_df_fig}
\end{figure}

Let us proceed with the derivation of the constraints emerging from the SC. Using the equations (\ref{criterion1}), (\ref{criterion2}) and (\ref{criterion3}) and setting $N=60$ and $\kappa=1$ we arrive at:

\begin{equation}
    \centering\label{E1_sc1}
    \Delta\phi= \sqrt{\frac{3}{2}} \sqrt{w} \left(\log \left(1+\frac{2 \sqrt{\beta(\gamma,\alpha)}}{\sqrt{3}\sqrt{w}}\right)-\log \left(1+\frac{2 \sqrt{\beta(\gamma,\alpha)}}{\sqrt{3} \sqrt{w}}+\frac{80\beta(\gamma,\alpha)}{w}\right)\right) ,
\end{equation}

\begin{equation}
    \centering\label{E1_sc2}
   \frac{|V'(\phi_i)|}{\kappa V(\phi_i)} = \frac{\sqrt{6}\sqrt{w}}{\sqrt{3}\sqrt{w}\sqrt{\beta(\gamma,\alpha)}+120\beta(\gamma,\alpha)},
\end{equation}

\begin{equation}
    \centering\label{E1_sc3}
  - \frac{V''(\phi_i)}{\kappa^2 V(\phi_i)} = \frac{-3w +2\sqrt{3}\sqrt{w}\sqrt{\beta(\gamma,\alpha)}+240\beta(\gamma,\alpha)} {\left(\sqrt{3}\sqrt{w}+120\sqrt{\beta}\right)^2\beta(\gamma,\alpha)}.
\end{equation}
Taking the relations (\ref{E1_sc1}), (\ref{E1_sc2}) and (\ref{E1_sc3}) we get Fig.\ref{E1_sc_desitter_fig} and Fig.\ref{E1_df_fig}.

\begin{figure}[h]
\centering \label{Comb.E}
\includegraphics[width=18pc]{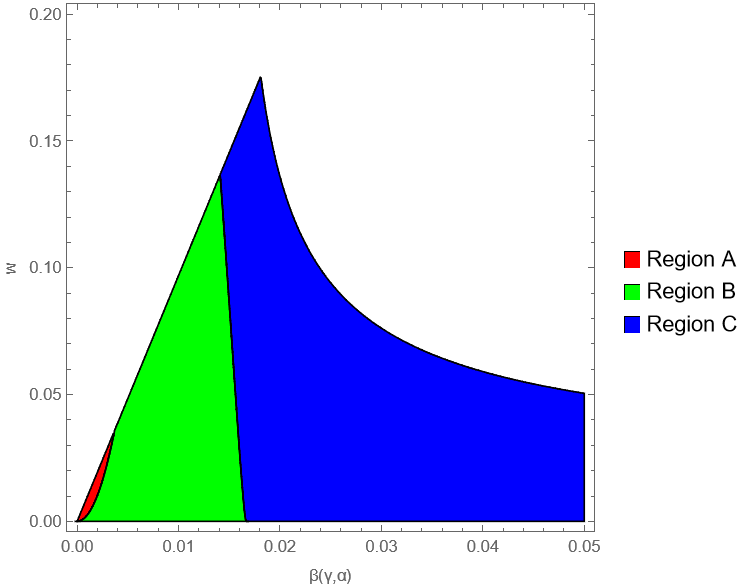}
\caption{Region plot for the combined constraints for E Model(n=1). In all highlighted Regions the Planck Constraints and at least one Swampland Criterion are satisfied. Region A is the area where all the SC are satisfied, in Region B (\ref{criterion1}) and (\ref{criterion3}) are satisfied simultaneously. In Region C (\ref{criterion1}) is satisfied.}
\label{E_Combin.fig}
\end{figure}

For (\ref{criterion1}) we get,
\begin{equation}
    \centering \label{E-MODEL-SW1}
    \beta(\gamma,\alpha)< \left[\frac{\sqrt{w}}{80 \sqrt{3}} \left( \exp{\left( \sqrt{\frac{2}{3 w}}\right)}-1 \right)+\frac{1}{480} \sqrt{2880\left( \exp{\left( \sqrt{\frac{2}{3 w}}\right)}-1 \right)w+12 w \left( \exp{\left( \sqrt{\frac{2}{3 w}}\right)}-1 \right)^2} \right]^2
\end{equation}
For the de Sitter conjecture to be satisfied, we derived the precise forms of the constraints for SC and conclude at,
\begin{equation}
    \centering \label{E-MODEL-SW2}
    w > \frac{13576.5 \sqrt{\beta^5(\gamma,\alpha)}+4800\beta^2(\gamma,\alpha)+9600 \beta^2(\gamma,\alpha)}{\left(\beta(\gamma,\alpha)-2 \right)^2}.
\end{equation}
Afterwards, we used the third SC and obtained the following inequalities,
\begin{equation}
\centering \label{E-MODEL-SW3a}
    \begin{aligned}
        w\leq & \frac{0.666667}{(\beta(\gamma,\alpha) +1)^2} \left(121 \beta(\gamma,\alpha)-7320\beta^2(\gamma,\alpha)+7200 \beta^3(\gamma,\alpha)+ \right. \\
        & \left.+\sqrt{241 \beta^2(\gamma,\alpha) - 72240 \beta^3(\gamma,\alpha) + 6.9264\cdot 10^6 \beta^4(\gamma,\alpha) - 2.0736 \cdot 10^8 \beta^5(\gamma,\alpha)} \right),
    \end{aligned}
\end{equation}
for $\beta(\gamma,\alpha)<0.00833333$, while for $0.00833333<\beta(\gamma,\alpha)<0.0166667$ we obtained,
\begin{equation}
\centering \label{E-MODEL-SW3b}
    \begin{aligned}
        w\leq & \frac{0.666667}{(\beta(\gamma,\alpha) +1)^2} \left(121 \beta(\gamma,\alpha)-7320\beta^2(\gamma,\alpha)+7200 \beta^3(\gamma,\alpha)- \right. \\
        & \left.-\sqrt{241 \beta^2(\gamma,\alpha) - 72240 \beta^3(\gamma,\alpha) + 6.9264\cdot 10^6 \beta^4(\gamma,\alpha) - 2.0736 \cdot 10^8 \beta^5(\gamma,\alpha)} \right),
    \end{aligned}
\end{equation}
We present all the mentioned results in FIG. \ref{E_Combin.fig}. We note that Region C of FIG. \ref{E_Combin.fig}. extents to $\beta(\gamma,\alpha)=1$ which contains GR, for $\gamma=1$ and $\alpha=0$

\subsection{T-Model(m=1) Model}\label{T1 model}

We proceed this work with the T-Model (m=1):
\begin{equation}
\centering \label{t_model_potential}
V(\phi)=\Lambda ^4 \tanh ^2\left(\frac{\phi}{\frac{\sqrt{6 w}}{\kappa}}\right),
\end{equation}
where $\Lambda$ has dimension mass dimension, so $[\Lambda]=m$ and $w$ has no dimensions following the general philosophy to work with dimensionless parameters. By following the methodology provided in \ref{methodology} we find $\phi_i$:

\begin{equation}
\centering\label{T1_fini}
\phi_i=\sqrt{\frac{3}{2}} \sqrt{w} \cosh ^{-1}\left(\frac{3 w \sqrt{\frac{4 \beta(\gamma,\alpha) }{3 w}+1}+4 \beta(\gamma,\alpha)  N}{3 w}\right).
\end{equation}
Following the methodology presented in Section \ref{methodology} and setting $\kappa=1$ we arrived at:

\begin{equation}
\centering \label{T-1_ns}
n_s= 1-\frac{4 \beta(\gamma,\alpha) \operatorname{csch}\left( \frac{1}{2} \cosh^{-1}\left(\frac{80\beta(\gamma,\alpha)}{w}+\sqrt{1+\frac{4 \beta(\gamma,\alpha)}{3 w}}\right)\right)}{3 w},
\end{equation}
and
\begin{equation}
\centering \label{T-1_r}
r= \frac{64\beta(\gamma,\alpha) \operatorname{csch}\left(\frac{1}{2}\cosh^{-1}{\left(\frac{80\beta(\gamma,\alpha)}{w}+\sqrt{1+\frac{4 \beta(\gamma,\alpha)}{3 w}}\right)}\right)}{3 w} .
\end{equation}

\begin{figure}[h]
\centering
\includegraphics[width=18pc]{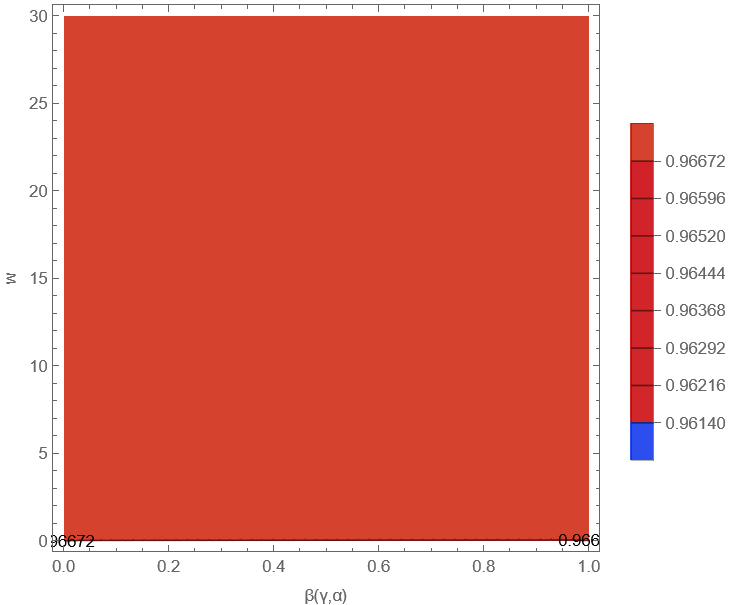}
\includegraphics[width=18pc]{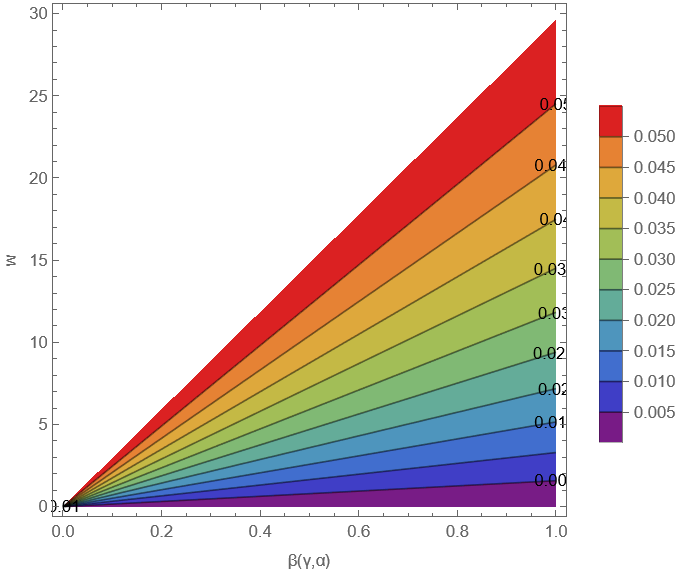}
\caption{Contour plot for the spectral index of primordial scalar curvature perturbations $n_s$ (left plot) and the tensor-to-scalar ratio $r$ (right plot) for $\beta(\gamma,\alpha)=[0, 1]$, $w=[10^{-6}, 30]$ and $N= 60$ for the T-Model (m=1)}\label{T1_INFL_fig}
\end{figure}

Using the above equations the demanded plots can be found to obtain the constraints, which are presented in the FIG.\ref{T1_INFL_fig} and it makes obvious that there is indeed an area that constraints the free parameters in such a way that the observational conditions are met. Using (\ref{PlankConstraints}) and combine them with (\ref{T-1_ns}) and (\ref{T-1_r}) one can compute the inequality,
\begin{equation}
\centering\label{T-nsandr}
\frac{\beta(\gamma,\alpha)}{w}\geq 0.033759,
\end{equation}
which satisfy both $n_s$ and $r$ indices, so the model is viable.

\begin{figure}
\centering
\includegraphics[width=18pc]{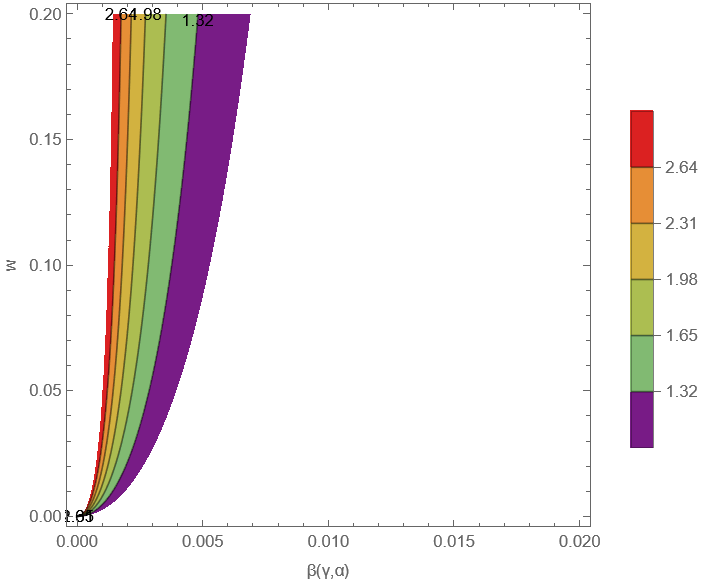}
\includegraphics[width=18pc]{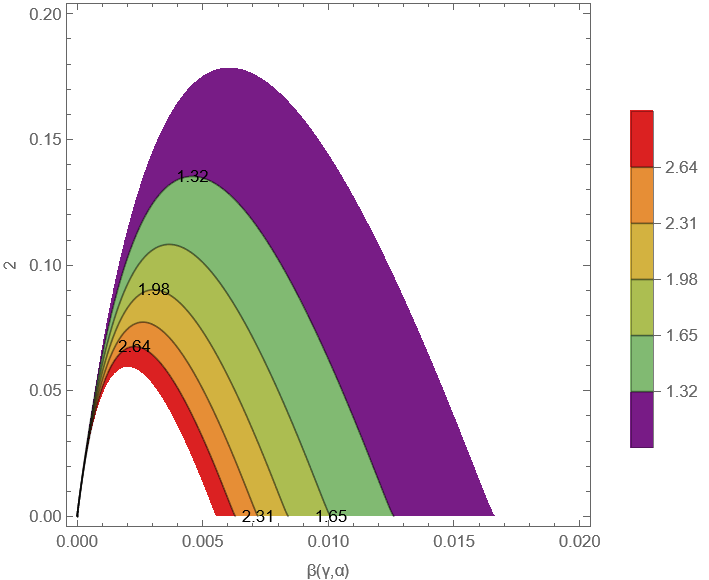}
\caption{Contour plot for de Sitter conjecture. The left plot provides $V'(\phi_i)/V(\phi_i)$ and the right plot provides the $-V''(\phi_i)/V(\phi_i)$. For both cases the constraints from SC are taken into account. Note that $\beta(\gamma,\alpha)=[0, 0.02]$, $w=[0, 0.2]$ (left plot) and $\beta(\gamma,\alpha)=[0, 0.02]$, $w=[0, 0.2]$ (left plot) and $N= 60$ for T-model (m=1).}\label{T1_sc_desitter_fig}
\end{figure}

\begin{figure}
\centering
\includegraphics[width=18pc]{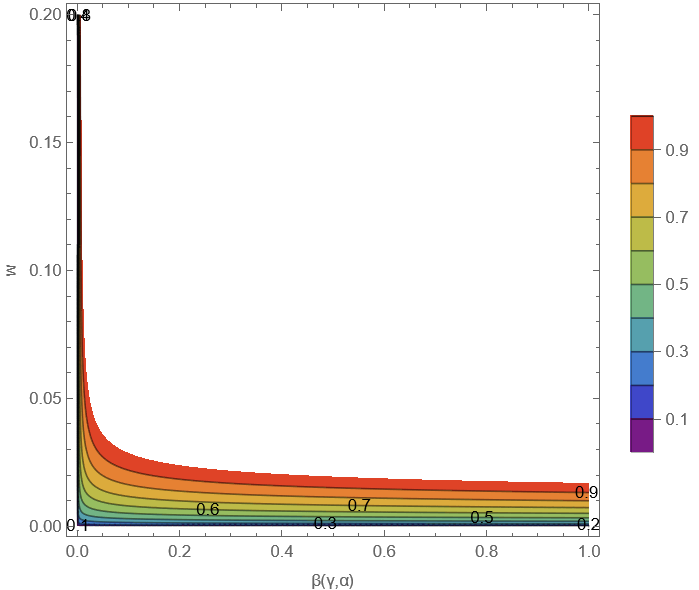}
\caption{Contour plot for the Swampland distance conjecture. Note that $\beta(\gamma,\alpha)=[0, 1]$, $w=[0, 0.2]$ and $N= 60$ for T-model (m=1).}\label{T1_df_fig}
\end{figure}

Let us proceed with the derivation of the constraints emerging from the SC. Using the equations (\ref{criterion1}), (\ref{criterion2}) and (\ref{criterion3}) and setting $N=60$ and $\kappa=1$ we arrive at:

\begin{equation}
    \centering\label{T1_sc1}
    \Delta\phi= \sqrt{\frac{3}{2}} \sqrt{w} \left[\sinh ^{-1}{\left(\frac{2 \sqrt{\beta(\gamma,\alpha)}}{\sqrt{3 w}}\right)}-\cosh ^{-1}{\left(\frac{80 \beta(\gamma,\alpha)}{w}+\sqrt{1+\frac{4 \beta(\gamma,\alpha)}{3 w}}\right)}\right] ,
\end{equation}

\begin{equation}
    \centering\label{T1_sc2}
   \frac{|V'(\phi_i)|}{\kappa V(\phi_i)} = \frac{2 \sqrt{6}\sqrt{w}}{\sqrt{\left[240+w \left(\sqrt{9+\frac{12 \beta(\gamma,\alpha)}{w}}-3\right)\right]\left[240+w \left(\sqrt{9+\frac{12 \beta(\gamma,\alpha)}{w}}+3\right)\right]}},
\end{equation}

\begin{equation}
    \centering\label{T1_sc3}
  - \frac{V''(\phi_i)}{\kappa^2 V(\phi_i)} = \frac{240 \beta(\gamma,\alpha)+w\left( \sqrt{9+\frac{12\beta(\gamma,\alpha)}{w}}-6\right)}{3 \beta(\gamma,\alpha)\left( w+4800\beta(\gamma,\alpha)+40 w\sqrt{9+\frac{12\beta(\gamma,\alpha)}{w}}\right) }.
\end{equation}
Taking the equations (\ref{T1_sc1}), (\ref{T1_sc2}) and (\ref{T1_sc3}) we will arrive at the figures Fig. \ref{T1_sc_desitter_fig} and Fig.\ref{T1_df_fig}.

Though simplifying the SC inequalities analytically is challenging we can use numerical methods to get the approximate form of those inequalities in the interval $\beta(\gamma,\alpha) \in [0,0.3]$ Eventually, we calculate that for (\ref{criterion1}),
\begin{equation}
    \centering\label{CR1-T}
    \begin{aligned}
    w\leq & 0.359627-\frac{0.134344}{\beta(\gamma,\alpha)}+\frac{0.0194468}{\beta^2(\gamma,\alpha)}-\frac{0.00139504}{\beta^3(\gamma,\alpha)}+\frac{0.0000577905}{\beta^4(\gamma,\alpha)}-\frac{1.50205\cdot 10^{-6}}{\beta^5(\gamma,\alpha)}+\frac{2.57041\cdot 10^{-8}}{\beta^6(\gamma,\alpha)}- \\
    & -\frac{2.96942 \cdot 10^{-10}}{\beta^7(\gamma,\alpha)}+\frac{2.32938 \cdot 10^{-12}}{\beta^8(\gamma,\alpha)}-\frac{1.22287 \cdot 10^{-14}}{\beta^9(\gamma,\alpha)}+\frac{4.11231 \cdot 10^{-17}} {\beta^{10}(\gamma,\alpha)}-\frac{8.00778\cdot 10^{-20}}{\beta^{11}(\gamma,\alpha)}+\frac{6.86655\cdot 10^{-23}}{\beta^{12}(\gamma,\alpha)}.
    \end{aligned}
\end{equation}
\\As for (\ref{criterion2}) we obtain,
\begin{equation}
    \centering\label{CR2-T}
    w \geq -0.113041 + 166.212 \beta(\gamma,\alpha) - 73406.1 \beta^{2}(\gamma,\alpha) + 1.49924\cdot 10^7 \beta^{3}(\gamma,\alpha) -
 1.30978 \cdot 10^9 \beta^{4}(\gamma,\alpha) + 4.51193*10^{10} \beta^{5}(\gamma,\alpha),
\end{equation}
Finally, for (\ref{criterion3}) we calculate,
\begin{equation}
    \centering\label{CR3-T}
    \begin{aligned}
    w\leq & 0.376333- 8.10773 \cdot 10^{-50} \beta^{-13}(\gamma,\alpha) + 8.20057 \cdot 10^{-45}\beta^{-12}(\gamma,\alpha) - 3.64875 \cdot 10^{-40} \beta^{-11}(\gamma,\alpha) +\\
    & + 9.42578 \cdot 10^{-36} \beta^{-10}(\gamma,\alpha) - 1.57122 \cdot 10^{-31} \beta^{-9}(\gamma,\alpha) + 1.77654 \cdot 10^{-27} \beta^{-8}(\gamma,\alpha) - 1.39423 \cdot 10^{-23} \beta^{-7}(\gamma,\alpha) +\\
    & +7.64346 \cdot 10^{-20} \beta^{-6}(\gamma,\alpha) - 2.90885 \cdot 10^{-16} \beta^{-5}(\gamma,\alpha) + 7.55089 \cdot 10^{-13} \beta^{-4}(\gamma,\alpha) - 1.2993 \cdot 10^{-9} \beta^{-3}(\gamma,\alpha) +\\ & +1.42555\cdot 10^-6 \beta^{-2}(\gamma,\alpha) - 0.000953931\beta^{-1}(\gamma,\alpha) - 8.0802 \beta(\gamma,\alpha) - 697.181 \beta^2(\gamma,\alpha) - 0.21229 \beta^3(\gamma,\alpha) +\\
    & +0.129271 \cdot 10^{-3} \beta^4(\gamma,\alpha).
    \end{aligned}
\end{equation}
It should be noted that the two above approximation fail very close to $\beta(\gamma,\alpha)=0$.

\begin{figure}[h]
\centering
\includegraphics[width=18pc]{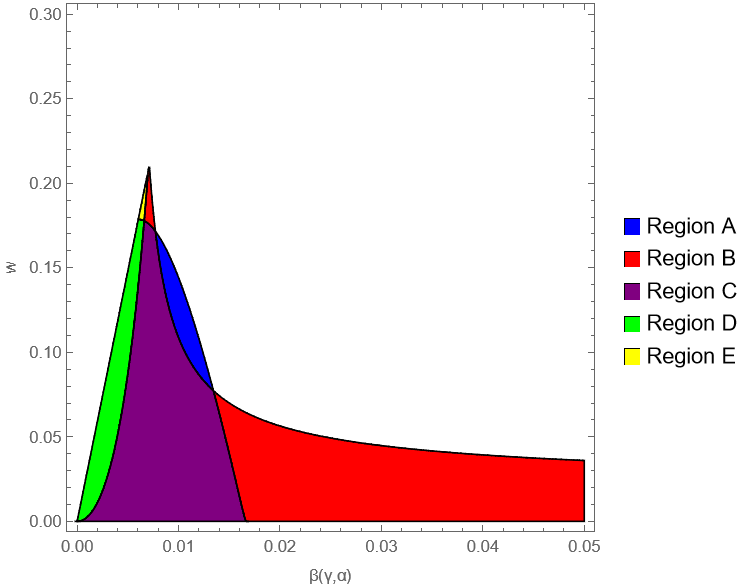}
\caption{Region plot for the combined constraints for T-Model(m=1). In all highlighted Regions the Planck Constraints and at least one Swampland Criterion are satisfied. In Region A (\ref{criterion3}) is satisfied and in Region B (\ref{criterion1}) is satisfied. In Region C (\ref{criterion1}) and (\ref{criterion3}) are satisfied simultaneously. Region D is the area where all the SC are satisfied. In Region E (\ref{criterion1}) and (\ref{criterion2}) are satisfied simultaneously.}
\label{T_Combin.fig}
\end{figure}

We present the combination of the Planck constraints and the  Swampland Criteria in FIG. \ref{T_Combin.fig}, we should mention that Region B of FIG. \ref{T_Combin.fig}. extents to $\beta(\gamma,\alpha)=1$ which contains GR, for $\gamma=1$ and $\alpha=0$.

\subsection{Modular Potential}\label{Modular Potential}
In this case, we are going to study the following potential:
\begin{equation}
    \centering\label{exp diko toy potential}
    V(\phi)=\Lambda^4 \left(1-d \exp(-w\kappa \phi) \right),
\end{equation}
where $\Lambda$ has dimensions of mass so $[\Lambda]=m$, $\lambda$ is dimensionless and $d$ has the following dimensions $[d]=m^4$. By following the methodology provided in \ref{methodology} we find $\phi_i$:

\begin{equation}
\centering\label{M_fini}
\phi_i= \frac{\log \left(\frac{1}{2} d \left(2 \beta(\gamma,\alpha)  w^2 N+\sqrt{2} \sqrt{\beta(\gamma,\alpha) } w+2\right)\right)}{\kappa  w}.
\end{equation}
So using the methodology presented in Section \ref{methodology} and setting $\kappa=1$ we arrived at:

\begin{equation}
\centering \label{M_ns}
n_s= 1-\frac{4 w \sqrt{\beta(\gamma,\alpha)}}{\sqrt{2}+2w N \sqrt{\beta(\gamma,\alpha)}}-\frac{12}{\left(\sqrt{2}+2w N \sqrt{\beta(\gamma,\alpha)}\right)^2},
\end{equation}
and
\begin{equation}
\centering \label{M_r}
r= \frac{32}{\left(\sqrt{2}+2w N \sqrt{\beta(\gamma,\alpha)}\right)^2} .
\end{equation}

\begin{figure}
\centering
\includegraphics[width=18pc]{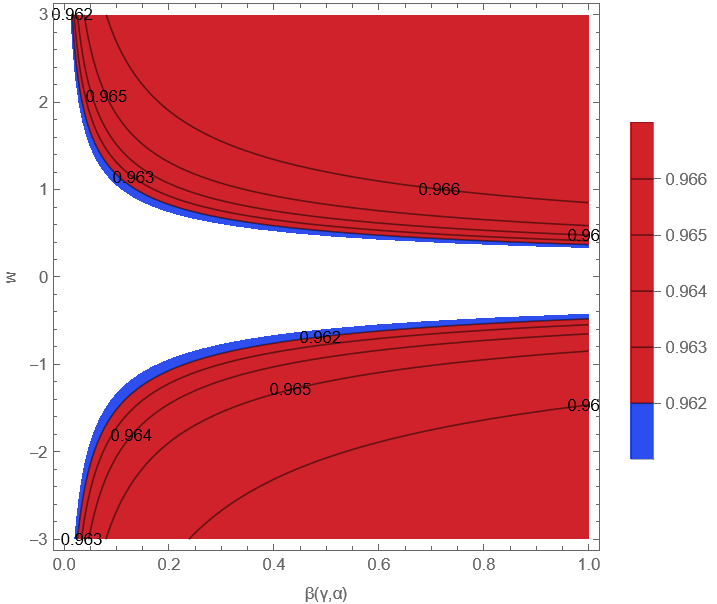}
\includegraphics[width=18pc]{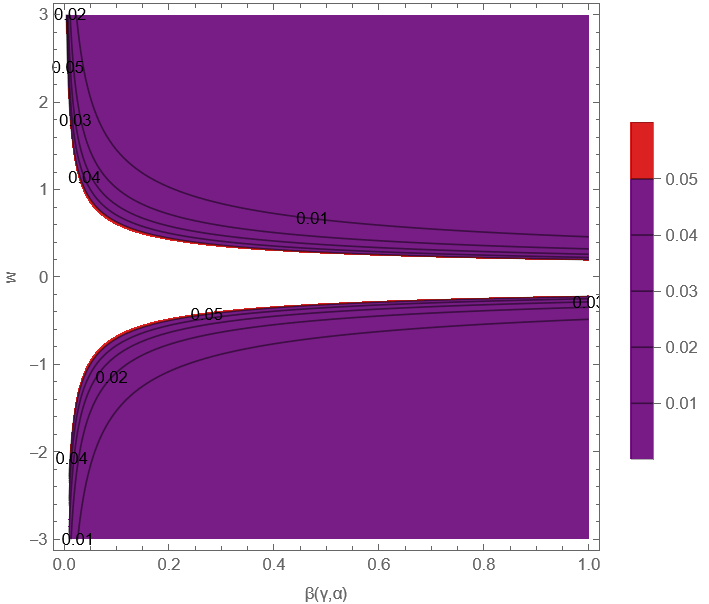}
\caption{Contour plot for the spectral index of primordial scalar curvature perturbations $n_s$ (left plot) and the tensor-to-scalar ratio $r$ (right plot) for $\beta(\gamma,\alpha)=[0, 1]$, $w=[-3, 3]$ and $N= 60$ for the Modular Potential.}\label{M_INFL_fig}
\end{figure}

Using the above equations the demanded plots can be found to obtain the constraints, which are presented in the Fig.\ref{M_INFL_fig} and it makes obvious that there is indeed an area that constraints the free parameters in such a way that the observational conditions are met. Taking the Plank Constraints (\ref{PlankConstraints}) into account, and put them all together, we calculate
\begin{equation}
\centering \label{MP_ns}
 \left\{
\begin{array}{ll}
        \beta(\gamma,\alpha) w^2\geq 0.109204 \, , \,  w>0 \\
        \beta(\gamma,\alpha) w^2\geq 0.17629  \, , \,  w<0
\end{array}
\right.
\end{equation}
for $n_s$ and \begin{equation}
\centering \label{MP_r}
\left\{
\begin{array}{ll}
      \beta(\gamma,\alpha) w^2\geq 0.0351261 \, , \,  w>0 \\
      \beta(\gamma,\alpha) w^2\geq 0.0445167  \, , \,  w<0
\end{array}
\right.
\end{equation}
for r.
As a result, combining (\ref{MP_ns}) and (\ref{MP_r}) both, it can be seen that for
\begin{equation}
\centering \label{MP_ns&r}
\left\{
\begin{array}{ll}
      \beta(\gamma,\alpha) w^2\geq 0.0351261 \, , \,  w>0 \\
      \beta(\gamma,\alpha) w^2\geq 0.0445167  \, , \,  w<0
\end{array}
\right.
\end{equation}
the model is viable.

Let us proceed with the derivation of the constraints emerging from the SC. Using the equations (\ref{criterion1}), (\ref{criterion2}) and (\ref{criterion3}) and setting $N=60$ and $\kappa=1$ we arrive at:

\begin{equation}
    \centering\label{M_sc1}
    \Delta\phi= \frac{\log \left(\frac{\sqrt{\beta(\gamma,\alpha)} d w}{\sqrt{2}}+d\right)-\log \left(d+\frac{\sqrt{\beta(\gamma,\alpha)} d w}{\sqrt{2}}+60 w^2 \beta(\gamma,\alpha) d\right)}{w} ,
\end{equation}

\begin{equation}
    \centering\label{M_sc2}
   \frac{|V'(\phi_i)|}{\kappa V(\phi_i)} = \frac{2}{\sqrt{2 \beta(\gamma,\alpha)}+120w \beta(\gamma,\alpha)},
\end{equation}

\begin{equation}
    \centering\label{M_sc3}
  - \frac{V''(\phi_i)}{\kappa V(\phi_i)} = \frac{2 w}{\sqrt{2 \beta(\gamma,\alpha)}+120 w \beta(\gamma,\alpha)}.
\end{equation}

\begin{figure}
\centering
\includegraphics[width=18pc]{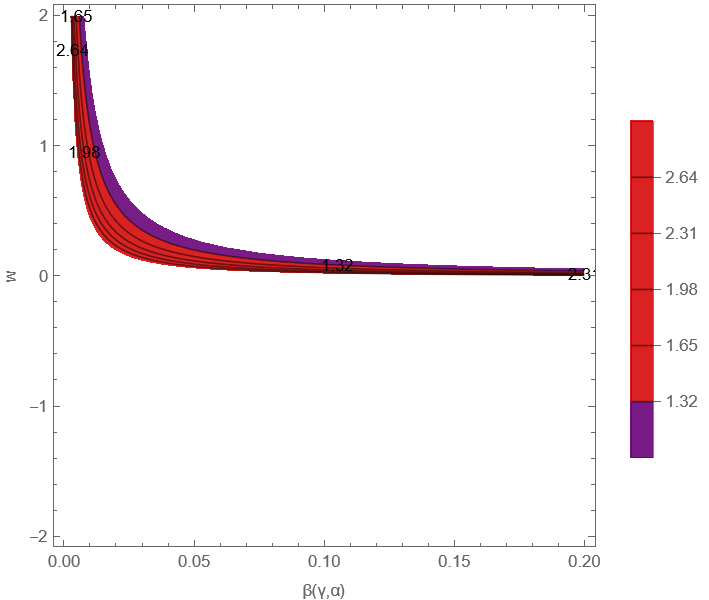}
\includegraphics[width=18pc]{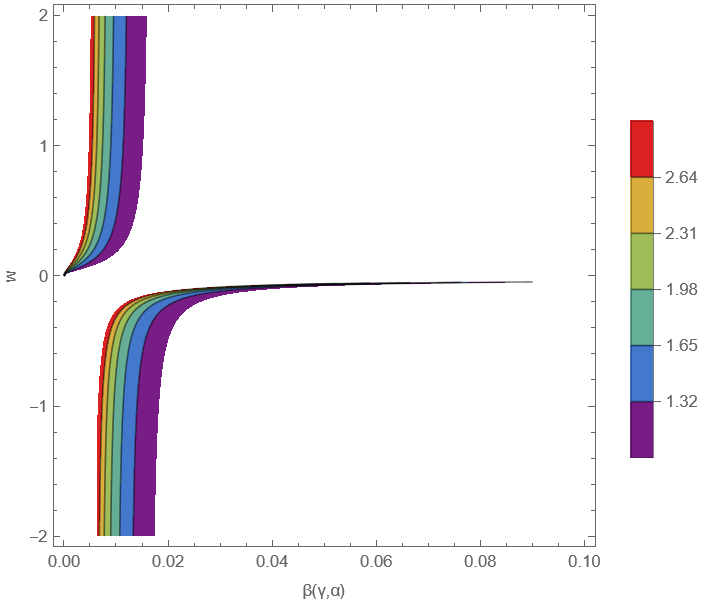}
\caption{Contour plot for de Sitter conjecture. The left plot provides $V'(\phi_i)/V(\phi_i)$ and the right plot provides the $-V''(\phi_i)/V(\phi_i)$. For both cases, the constraints from the SC are taken into account. Note that $\beta(\gamma,\alpha)=[0, 0.2]$, $w=[-2, 2]$ (left plot) and $\beta(\gamma,\alpha)=[0, 0.1]$, $w=[-2, 2]$ (left plot) and $N= 60$ for modular potential.}\label{M_sc_desitter_fig}
\end{figure}

\begin{figure}
\centering
\includegraphics[width=18pc]{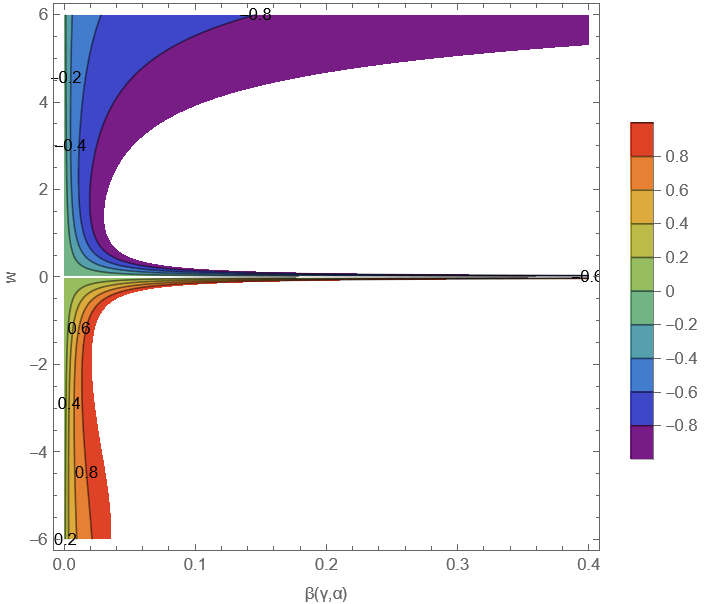}
\caption{Contour plot for the Swampland distance conjecture. Note that $\beta(\gamma,\alpha)=[0, 0.4]$, $w=[-6,6]$ and $N= 60$ for a modular potential.}\label{M_df_fig}
\end{figure}

Taking the equations (\ref{M_sc1}), (\ref{M_sc2}) and (\ref{M_sc3}) we will arrive at the figures Fig. \ref{M_sc_desitter_fig} and Fig. \ref{M_df_fig}.

It is apparent that the expression (\ref{M_sc1}) is very complicated. Therefore, the inequality of the first SC (\ref{criterion1}) cannot be simplified. However, the combination of (\ref{criterion2}) and (\ref{M_sc2}) can be simplified further,
\begin{equation}
     \centering\label{M_SC2}
     -\frac{1}{60 \sqrt{2\beta(\gamma,\alpha)}}< w\leq \frac{1}{60 \beta(\gamma,\alpha)}-\frac{1}{60 \sqrt{2\beta(\gamma,\alpha)}}
\end{equation}

Furthermore, combining the (\ref{criterion3}) as well as (\ref{M_sc3}) we calculated the following expressions,
\begin{equation}{
\centering\label{M_SC3}
    \left\{
    \begin{array}{ll}
      \frac{1}{7200 w^2} < \beta(\gamma,\alpha)\leq \frac{1+240 w^2}{14400}+\frac{\sqrt{1+480 w^2}}{14400 w^2},\, \ w<0

      \\ \\ \beta(\gamma,\alpha) \leq \frac{1+240 w^2}{14400 w^2}-\sqrt{\frac{1+480w^2}{w^4}},
      \, 0<w<\frac{1}{2\sqrt{15}}

      \\ \\ \beta(\gamma,\alpha) \leq \frac{1}{120}, \, \ w=\frac{1}{2\sqrt{15}}

      \\ \\ \beta(\gamma,\alpha) \geq \frac{1+240 w^2}{14400 w^2}-\sqrt{\frac{1+480w^2}{w^4}},
      \, w>\frac{1}{2\sqrt{15}} .
\end{array}
 \right. }
\end{equation}

The combination of the Planck constraints and the  Swampland Criteria is presented in FIG. \ref{T_Combin.fig}, we should mention that Region C of FIG. \ref{T_Combin.fig}. extents to $\beta(\gamma,\alpha)=1$ which contains GR, for $\gamma=1$ and $\alpha=0$.

\begin{figure}[h]
\centering
\includegraphics[width=18pc]{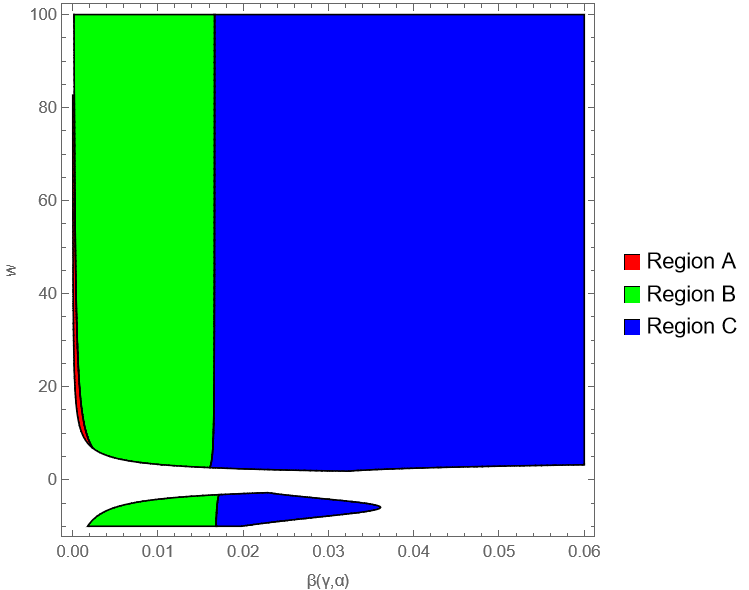}
\caption{Region plot for the combined constraints for Modular Potential. In all highlighted Regions the Planck Constraints and at least one Swampland Criterion are satisfied. Region A is the Region where all the Swampland criteria are satisfied. In Region B (\ref{criterion1}) and (\ref{criterion3}) are satisfied simultaneously. In Region C (\ref{criterion1}) is satisfied.}
\label{Modular_Combin.fig}
\end{figure}

\section{Conclusions}
In this work, we considered the $f(R,T)=\gamma R+2\kappa \alpha T$ gravity with a canonical minimally coupled scalar field as a model for inflation, with $\alpha \geq 0$ and $\gamma$. The main motivation behind this study was that the mentioned $f(R,T)$ gravity yields similar relations for the slow-roll indices as a rescaled Einstein-Hilbert gravity.
As we demonstrated, not all potentials generate viable results, more specifically, the Power law potentials proved to be non-viable, as the tensor-to-scalar ratio and the scalar spectral index are not compatible with the Planck Constraints in the same region of the parameter space.  However, the rest of the studied models are compatible with both the latest Planck data (2018) and the Swampland Criteria. Using both analytical and numerical tools we were able to specify the region of the parameter space where both the Planck Constraints and the SC are satisfied and we presented the relevant region plots. Unfortunately, because $\beta(\gamma,\alpha)$ must be positive, the tensor spectral index can only be negative, which means that the primordial gravitational waves produced will not be observed by any of the next-generation gravitational wave detectors. An important finding of this work is the simultaneous satisfaction of the Swampland Criteria and the Planck constraints in the GR limit. In future work, we would like to investigate the non-minimally coupled case, as a further generalization of this work.

\end{document}